# Solution of nonlinear space–time fractional differential equations via the fractional projective Riccati expansion method

Emad A-B. Abdel-Salam[a, d], Eltayeb A. Yousif[b, d] and Gmal F. Hassan[c, d]

a: Mathematics Department, Faculty of Science, Assiut University, New Valley Branch, El-Kharja 72511, Egypt. Email:emad_abdelsalam@yahoo.com.

b: Applied Mathematics Department, Faculty of Mathematical Sciences, University of Khartoum, 11111 Khartoum, Sudan

c: Mathematics Department, Faculty of Science, Assiut University, Assiut 71516, Egypt

d: Mathematics Department, Faculty of Science, Northern Border University, Arar 91431, Saudi Arabia.

**Abstract.**

In this paper, the fractional projective Riccati expansion method is proposed to solve fractional differential equations. To illustrate the effectiveness of the method, we discuss the space-time fractional Burgers equation, the space-time fractional mKdV equation and time fractional biological population model. The solutions are expressed in terms of fractional hyperbolic functions. These solutions are useful to understand the mechanisms of the complicated nonlinear physical phenomena and fractional differential equations. Among these solutions, some are found for the first time. The fractal index for the obtained results is equal to one. Counter examples to compute the fractal index are introduced in appendix.

## 1 Introduction

Nowadays, the fractional calculus is successfully used to investigate many complex nonlinear phenomena. It has many important applications in different research areas and engineering applications. At the same time, the fractional differential equations (FDEs) become more and more popular for describing systems identification. Examples include: fluid flow, control problem, signal processing, viscoelastic materials, polymers, fluid mechanics, finance, biology, physics, engineering and other areas of science. The fractional order partial differentiation is a generalization of the classical integer order partial differentiation. Fractional derivatives provide an excellent instrument for the description of memory properties of various processes [1- 47]. This is the main advantage of fractional derivatives in comparison with classical integer- order models, in which such effects are in fact neglected [1- 7]. Searching for analytical and numerical solutions of FDEs is currently a very active area of research. In the past two decades, both mathematicians and physicists have made much significant work in this direction and presented some effective methods. Examples include: Laplace transform, Fourier transform, finite difference, finite element, Adomian decomposition, differential transform, variational iteration, homotopy perturbation, exp-function, $(G'/G)$-expansion, fractional sub-equation, generalized fractional sub-equation, fractional Riccati expansion, improved fractional Riccati expansion and fractional mapping [8- 20].

The projective Riccati expansion method first introduced by Conte et al. to seek more new solitary wave solutions of some nonlinear partial differential equations (NLPDEs) of integer orders that can be expressed as a polynomial in two elementary functions which satisfy a project Riccati equations [21]. Yan developed Conte's method and presented the general projective Riccati equation method [22]. Many authors successfully applied the projective Riccati expansion method to get new solutions of NLPDEs [23- 29]. Moreover, El-Sabbagh et al extended the projective Riccati expansion method and get new solutions to NLPDEs by introducing the two deformation parameters $p$ and $q$ in the definition of the hyperbolic and trigonometric functions. In this paper, we introduce the fractional projective Riccati expansion method to solve FDEs with the modified Riemann-Liouville derivative defined by Jumarie [30- 34].

This paper organized as follows: brief introduction of the fractional calculus and the description of the fractional projective Riccati expansion method are introduced in section 2. In section 3, the solution of the space-time fractional Burgers equation, the space-time fractional mKdV equation and time fractional biological population model are studied. In section 4, discussion and conclusion are presented.

**2. Preliminaries and Fractional projective Riccati expansion method**

Fractional Calculus is the generalizations of the classical calculus. It provides a redefinition of mathematical tools and it is very useful to deal with anomalous and frictional systems [35- 47]. There are different kinds of fractional integration and differentiation operators. The most famous one is the Riemann-Liouville definition [1– 7], which has been used in various fields of science and engineering successfully, but this definition leads to the result that constant function differentiation is not zero. Caputo put definitions which give zero value for fractional differentiation of constant function, but these definitions require that the function should be smooth and differentiable [1– 7]. Recently, Jumarie derived definitions for the fractional integral and derivative called modified Riemann-Liouville [30– 34]. Some advantages can be cited to the modified Riemann-Liouville definition by Jumarie, first the derivative of a constant function equal to zero and second, we use it so much for differentiable as non-differentiable functions [37- 35]. The modified Riemann-Liouville fractional definitions are used effectively in many different problems [30– 34]. The Jumarie's modified Riemann–Liouville derivative of order $\alpha$ is defined by the expression

$$D_x^{\alpha} f(x) = \begin{cases} \dfrac{1}{\Gamma(1-\alpha)} \displaystyle\int_0^x (x-\xi)^{-\alpha-1}[f(\xi)-f(0)]d\xi, & \alpha < 0 \\ \dfrac{1}{\Gamma(1-\alpha)} \dfrac{d}{dx} \displaystyle\int_0^x (x-\xi)^{-\alpha}[f(\xi)-f(0)]d\xi, & 0<\alpha<1 \\ [f^{(\alpha-n)}(x)]^{(n)}, & n \le \alpha < n+1,\ n \ge 1. \end{cases} \tag{1}$$

Some useful formulas and results of Jumarie's modified Riemann–Liouville derivative were summarized in [31]. Four of them (which will be used in the following sections) are

$$D_x^{\alpha} x^{\gamma} = \frac{\Gamma(\gamma+1)}{\Gamma(\gamma+1-\alpha)} x^{\gamma-\alpha}, \qquad \gamma > 0, \tag{2}$$

$$D_x^{\alpha}[f(x)g(x)] = g(x)D_x^{\alpha} f(x) + f(x)D_x^{\alpha} g(x), \tag{3}$$

$$D_x^{\alpha} f[g(x)] = f_g^{'}[g(x)]D_x^{\alpha} g(x), \tag{4}$$

$$D_x^{\alpha} f[g(x)] = D_g^{\alpha} f[g(x)](g_x^{'})^{\alpha}. \tag{5}$$

The above properties play an important role in the fractional projective Riccati expansion method. The formulas 3, 4 and 5 follow from the fractional Leibniz rule and the fractional Barrow's formula. In addition, Kolwankar obtained the same formula (3) by using an approach on Cantor space [36- 39]. Jumarie in [31] gave detailed proofs of the above formulas. Non-integer differentiability and randomness [35- 47] are mutually related in their nature, in such a way that studies on fractals on the one hand, and fractional Brownian motion on the other hand, are often parallel. The non-differentiable means that a function is continuous everywhere and it has a fractional derivative but do not have an integer-differentiable, necessarily exhibits random-like or pseudo-random features. This may explain huge amount of literature extending the theory of stochastic differential equations to describe stochastic dynamics driven by fractional Brownian motion.

We outline the main steps of the fractional projective Riccati expansion method for solving FDEs. For a given nonlinear FDE, say, in two variables $x$ and $t$

$$P(u,\ D_t^{\alpha}u, D_x^{\alpha}u,\ D_t^{2\alpha}u,\ D_x^{2\alpha}u,\ ...) = 0, \tag{6}$$

where $D_t^{\alpha}u$ and $D_x^{\alpha}u$ are Jumarie's modified Riemann–Liouville derivatives of $u$, $u = u(x, t)$ is unknown function, $P$ is a polynomial in $u$ and its various partial derivatives in which the highest order derivatives and nonlinear terms are involved.

**Step 1**. By using the travelling wave transformation:

$$u(x, t) = u(\xi), \qquad \xi = x + \omega t, \tag{7}$$

where $\omega$ is a constant to be determined later, the nonlinear FDE (5) is reduced to the following nonlinear fractional ordinary differential equation (FODE) for $u = u(\xi)$:

$$\tilde{P}(u,\ \omega^{\alpha} D_{\xi}^{\alpha}u, D_{\xi}^{\alpha}u,\ \omega^{2\alpha} D_{\xi}^{2\alpha}u,\ D_{\xi}^{2\alpha}u,\ ...) = 0, \tag{8}$$

**Step 2.** We assume that (8) has the following formal solution:

$$u(\xi) = a_0 + \sum_{i=1}^{n} f^{\,i-1}(\xi)\left[a_i\ f(\xi) + b_i\ g(\xi)\right], \tag{9}$$

with

$$D^\alpha f = -f\,g, \qquad D^\alpha g = 1 - g^2 - r\,f, \qquad g^2 = 1 - 2r\,f + (r^2 + \mu) f^2. \tag{10}$$

Using the Mittag-Leffler function in one parameter $E_\alpha(x) = \sum_{k=0}^{\infty} \frac{x^k}{\Gamma(1+\alpha k)} \quad (\alpha > 0)$, equation (10) has different solutions, so we ought to discuss solution of equation (10) in the following cases (detailed proof of these cases is in appendix A).

case 1: when $\mu = -M$, we have

$$f_1 = \frac{\operatorname{sech}(\xi,\alpha)}{r \operatorname{sech}(\xi,\alpha) + 1}, \qquad g_1 = \frac{\tanh(\xi,\alpha)}{r \operatorname{sech}(\xi,\alpha) + 1}, \tag{11}$$

$$f_2 = \frac{4}{5\cosh(\xi,\alpha) + 3\sinh(\xi,\alpha) + 4r}, \qquad g_2 = \frac{5\sinh(\xi,\alpha) + 3\cosh(\xi,\alpha)}{5\cosh(\xi,\alpha) + 3\sinh(\xi,\alpha) + 4r}, \tag{12}$$

case 2: when $\mu = M$, , we have

$$f_3 = \frac{\operatorname{csch}(\xi,\alpha)}{r \operatorname{csch}(\xi,\alpha) + 1}, \qquad g_3 = \frac{\coth(\xi,\alpha)}{r \operatorname{csch}(\xi,\alpha) + 1}, \tag{13}$$

where $M = E_\alpha(x^\alpha) E_\alpha(-x^\alpha)$, and $r$ is an arbitrary constant and the generalized hyperbolic and trigonometric functions are defined as

$$\cosh(\xi,\alpha) = \frac{E_\alpha(\xi^\alpha) + E_\alpha(-\xi^\alpha)}{2}, \quad \sinh(\xi,\alpha) = \frac{E_\alpha(\xi^\alpha) - E_\alpha(-\xi^\alpha)}{2},$$
$$\tanh(\xi,\alpha) = \frac{\sinh(\xi,\alpha)}{\cosh(\xi,\alpha)}, \qquad \coth(\xi,\alpha) = \frac{1}{\tanh(\xi,\alpha)},$$
$$\operatorname{sech}(\xi,\alpha) = \frac{1}{\cosh(\xi,\alpha)}, \qquad \operatorname{csch}(\xi,\alpha) = \frac{1}{\sinh(\xi,\alpha)}.$$

**Step 3**: We determine the positive integer $n$ in (9) by using the homogenous balance between the linear term of the highest order and the nonlinear term in equation (8). In some nonlinear deferential equations, the balance number $n$ is not a positive integer. In this case, we make the following transformations:

(a) When $n = \frac{q}{p}$ is a fraction, we let

$$u(\xi) = v^{\frac{q}{p}}(\xi), \tag{14}$$

then substituting (14) into (8) to get a new equation in the new function $v(\xi)$ with a positive integer balance number.

(b) When $n$ is negative number, we let

$$u(\xi) = v^{n}(\xi). \quad (15)$$

Substituting (15) into (8), we get a new equation in the new function $v(\xi)$ with a positive integer balance number.

**Step 4**: Substituting (9) with (10) into the FODE (8) and collecting coefficients of polynomials of $f^{i} g^{j}$, $(i = 0,1,....; j = 0,1)$, equating each coefficient to zero, we obtain a set of algebraic equations of $a_0, a_i, b_i$ $(i = 1, 2, ....., n)$ and $\omega$. Solving the algebraic system of equations to obtain $a_0, a_i, b_i$ and $\omega$. Choose properly $\mu$ and the functions $f$, $g$, substituting $a_0, a_i, b_i$, $\omega$ into (9) with (11) - (13), we have the formal solutions of (8).

## 4. Applications

***4.1*** *Space- time fractional Burgers equation*

The Burgers equation is an important nonlinear wave equation and this simple model explains diversity of nonlinear phenomena. The Burgers equation is an important diffusion equation in physics, which describe the far field of wave propagation in the corresponding dissipative system. It is arises in models of traffic and fluid flow. This equation possesses a fundamental quadratic nonlinearity and it is an appropriate starting model for studying turbulence. It is suitable to not only explore and validate numerical models but also serves as a reasonable means to study physical processes such as shock waves, acoustic transmission, traffic flow, supersonic flow around airfoils, waves under the influence of diffusion, wave propagation in a thermo-elastic medium and the dynamics in a liquid. The space- time fractional Burgers equation [15], which is the transformed generalization of the Burgers equation, is defined as follows:

$$D_t^{\alpha} u = \kappa u\, D_x^{\alpha} u + \tau\, D_x^{2\alpha} u, \quad (16)$$

where $\kappa$, $\tau$ are constants and $\alpha$ describing the order of the fractional derivatives $(0 < \alpha \le 1)$. In order to solve equation (16) by the fractional projective Riccati expansion method, we use the travelling wave transformation $u(x, t) = u(\xi)$, $\xi = x + \omega t$, where $\omega$ is the dimensionless velocity of the wave. Then, equation (16) reduced to the following nonlinear FODE:

$$\omega^{\alpha} D_{\xi}^{\alpha} u - \kappa\, u\, D_{\xi}^{\alpha} u - \tau\, D_{\xi}^{2\alpha} u = 0. \qquad (17)$$

By balancing $D_{\xi}^{2\alpha} u$ with $u\, D_{\xi}^{\alpha} u$ gives $n = 1$. Therefore, the solution of equation (16) can be expressed as

$$u(\xi) = a_0 + a_1\, f(\xi) + b_1\, g(\xi). \qquad (18)$$

Substituting (18) into (17) using (10) and setting the coefficients of $f^{i} g^{j}, (i = 0,1,2;$ $j = 0,1),$ to zero, we get

$$a_0 = \frac{\omega^{\alpha}}{\kappa}, \qquad a_1 = \pm \frac{\tau \sqrt{r^2 + \mu}}{2\kappa}, \qquad b_1 = \frac{\tau}{\kappa}. \qquad (19)$$

By selecting the special value of $\mu$ and the corresponding function $f(\xi),\ g(\xi)$, we get the following solutions of (16):

$$u_1(\xi) = \frac{\omega^{\alpha}}{\kappa} + \frac{\tau}{2\kappa}\left[\frac{2\tanh(\xi,\alpha) \pm \sqrt{r^2 - M}\ \operatorname{sech}(\xi,\alpha)}{r\operatorname{sech}(\xi,\alpha) + 1}\right], \qquad (20)$$

$$u_2(\xi) = \frac{\omega^{\alpha}}{\kappa} + \frac{\tau}{\kappa}\left[\frac{5\sinh(\xi,\alpha) + 3\cosh(\xi,\alpha) \pm 2\sqrt{r^2 - M}}{5\cosh(\xi,\alpha) + 3\sinh(\xi,\alpha) + 4r}\right], \qquad (21)$$

$$u_3(\xi) = \frac{\omega^{\alpha}}{\kappa} + \frac{\tau}{2\kappa}\left[\frac{2\coth(\xi,\alpha) \pm \sqrt{r^2 + M}\ \operatorname{csch}(\xi,\alpha)}{r\operatorname{csch}(\xi,\alpha) + 1}\right], \qquad (22)$$

where $\xi = x + \omega t.$ To understand the effect of the fractional order $\alpha$, we graph equations (20) and (21) with different values of $\alpha$ which represents a soliton solution. Figure (1-a), shows the solution (20) in 3-dimension when the values of the parameters $k = \tau = \omega = 1,\ r = 2$ when $\alpha = 0.75$. Figure (1- b), shows the relation of $u_1$ at different values of the fractional parameter $\alpha = 0.25,\ 0.5,\ 0.75,\ 1$. It has observed that the amplitude and the width of the wave increased as the values of the fractional order derivative increase. Figure (2- a), shows the solution (21) in 3-dimension when the values of the parameters $k = 2,\ \tau = \omega = 1,\ r = \sqrt{5}$ when $\alpha = 0.8$. Figure (1- b), shows the relation of $u_2$ at different values of the fractional parameter $\alpha = 0.25,\ 0.5,\ 0.75,\ 1$. It has observed that the amplitude decrease with the values of the fractional order derivative increase. Therefore, the fractional order can be used to modulate the soliton shape.

When $\alpha = 1$ equation (16) reduced to the well known Burgers equation

$$u_t = \kappa u u_x + \tau u_{xx}. \tag{23}$$

The solutions (20) - (22) take the form of the following solutions of the Burgers equation

$$u_{1bur}(\xi) = \frac{\omega}{\kappa} + \frac{\tau}{2\kappa}\left[\frac{2\tanh(\xi) \pm \sqrt{r^2 - 1}\,\mathrm{sech}(\xi)}{r\,\mathrm{sech}(\xi) + 1}\right], \tag{24}$$

$$u_{2bur}(\xi) = \frac{\omega}{\kappa} + \frac{\tau}{\kappa}\left[\frac{5\sinh(\xi) + 3\cosh(\xi) \pm 2\sqrt{r^2 - 1}}{5\cosh(\xi) + 3\sinh(\xi) + 4r}\right], \tag{25}$$

$$u_{3bur}(\xi) = \frac{\omega}{\kappa} + \frac{\tau}{2\kappa}\left[\frac{2\coth(\xi) \pm \sqrt{r^2 + 1}\,\mathrm{csch}(\xi)}{r\,\mathrm{csch}(\xi) + 1}\right], \tag{26}$$

***4.2*** *Space-time fractional mKdV equation*

The KdV equation is the earliest soliton equation that was firstly derived by Korteweg and de Vries to model the evolution of shallow water wave in 1895. The KdV-type equations have applications in shallow-water waves, optical solitons in the two cycle regime, density waves in traffic flow of two kinds of vehicles, short waves in nonlinear dispersive models, surface acoustic soliton in a system supporting long waves, quantum field theory, plasma physics and solid-state physics. The KdV type equations with the quadratic nonlinearity are important nonlinear models, which have been derived in many unrelated branches of sciences and engineering including the pulse-width modulation, mass transports in a chemical response theory, dust acoustic solitary structures in magnetized dusty plasmas and nonlinear long dynamo waves observed in the Sun. However, the higher-order nonlinear terms must be taken into account in some complicated situations like at the critical density or in the vicinity of the critical velocity. The modified KdV (mKdV) equation, has recently been discovered, e.g., to model the dust-ion-acoustic waves in such cosmic environments as those in the supernova shells and Saturn's F-ring. The space-time fractional mKdV equation [19], which is the transformed generalization of the mKdV equation, is defined as follows:

$$D_t^{\alpha} u + u^2 D_x^{\alpha} u + \tau D_x^{3\alpha} u = 0, \tag{27}$$

where $\tau$ is constant and $\alpha$ describing the order of the fractional derivatives $(0<\alpha\le 1)$. In order to solve equation (27) by using the fractional projective Riccati expansion method, we use the travelling wave transformation $u(x, t)=u(\xi),\ \xi=x+\omega t,$ where $\omega$ is the wave velocity. Then, equation (27) reduced to the following nonlinear FODE:

$$\omega^{\alpha} D_{\xi}^{\alpha}u+u^2 D_{\xi}^{\alpha}u+\tau D_{\xi}^{3\alpha}u=0. \tag{28}$$

By balancing $D_{\xi}^{3\alpha}u$ with $u^2 D_{\xi}^{\alpha}u$ gives $n=1$. Therefore, the solution of equation (27) can be expressed as

$$u(\xi)=a_0+a_1 f(\xi)+b_1 g(\xi). \tag{29}$$

Substituting (29) into (28) using (10) and setting the coefficients of $f^i g^j$, to zero, we obtain

$$\omega^{\alpha}=\frac{\tau}{2},\quad a_1=\frac{1}{2}\sqrt{-6(r^2+\mu)\tau},\qquad b_1=\frac{\sqrt{-6\tau}}{2},\quad a_0=0, \tag{30}$$

By selecting the special value of $\mu$ and the corresponding function $f(\xi),\ g(\xi)$, we get the following solutions of (26):

$$u_1=\frac{\sqrt{-6\tau}}{2}\left[\frac{\sqrt{r^2-M}\,\operatorname{sech}(\xi,\alpha)+\tanh(\xi,\alpha)}{r\operatorname{sech}(\xi,\alpha)+1}\right], \tag{31}$$

$$u_2=\frac{\sqrt{-6\tau}}{2}\left[\frac{4\sqrt{r^2-M}+5\sinh(\xi,\alpha)+3\cosh(\xi,\alpha)}{5\cosh(\xi,\alpha)+3\sinh(\xi,\alpha)+4r}\right], \tag{32}$$

$$u_3=\frac{\sqrt{-6\tau}}{2}\left[\frac{\sqrt{r^2+M}\,\operatorname{csch}(\xi,\alpha)+\coth(\xi,\alpha)}{r\operatorname{csch}(\xi,\alpha)+1}\right], \tag{33}$$

$\xi=x+\omega t,$ When $\alpha=1$ equation (26) reduced to the well known mKdV equation

$$u_t+u^2 u_x+\tau u_{xxx}=0. \tag{34}$$

The solutions (31) - (33) take the form of the following solutions of the mKdV equation

$$u_{1mkdv}=\frac{\sqrt{-6\tau}}{2}\left[\frac{\sqrt{r^2-1}\operatorname{sech}(\xi)+\tanh(\xi)}{r\operatorname{sech}(\xi)+1}\right], \tag{35}$$

$$u_{2mkdv}=\frac{\sqrt{-6\tau}}{2}\left[\frac{4\sqrt{r^2-1}+5\sinh(\xi)+3\cosh(\xi)}{5\cosh(\xi)+3\sinh(\xi)+4r}\right], \tag{36}$$

$$u_{3mkdv} = \frac{\sqrt{-6\tau}}{2}\left[\frac{\sqrt{r^2+1}\,\mathrm{csch}(\xi)+\coth(\xi)}{r\,\mathrm{csch}(\xi)+1}\right]. \tag{37}$$

*4.3 Time fractional biological population model*

The problem of biological diffusion is an issue of increasing significance in contemporary ecology. Mathematical aspects of the biological problem have been considered in many papers. The time fractional biological population model has the form

$$D_t^\alpha u = D_x^2 u + D_y^2 u + \tau(u^2-\sigma), \qquad 0<\alpha\le 1, \tag{38}$$

where $\tau, \sigma$ are constants, $u$ represents the population density and $\tau(u^2-\sigma)$ represents the population supply due to births and deaths. By using the travelling wave transformation $u(x, y, t) = u(\xi),\ \xi = k\,x + i\,k\,y + \omega t,\ i^2 = -1$ [18], equation (38) is reduced to the following nonlinear FODE:

$$\omega^\alpha D_\xi^\alpha u - \tau(u^2-\sigma) = 0. \tag{39}$$

Thus, the solution of equation (38) has the form

$$u(\xi) = a_0 + a_1\, f(\xi) + b_1\, g(\xi). \tag{40}$$

Substituting (40) into (39) using (10) and setting the coefficients of $f^i g^j$, to zero, we obtain

$$a_0 = 0, \quad \sigma = b_1^2, \quad \omega^\alpha = -2b_1\tau, \quad a_1 = b_1\sqrt{r^2+\mu} \tag{41}$$

By selecting the special value of $\mu$ and the corresponding function $f(\xi),\ g(\xi)$, we get the following solutions of (38):

$$u_1 = \frac{\sqrt{\sigma(r^2-M)}\,\mathrm{sech}(\xi,\alpha)+\sqrt{\sigma}\,\tanh(\xi,\alpha)}{r\,\mathrm{sech}(\xi,\alpha)+1}, \tag{42}$$

$$u_2 = \frac{4\sqrt{\sigma(r^2-M)}+\sqrt{\sigma}\left[5\sinh(\xi,\alpha)+3\cosh(\xi,\alpha)\right]}{5\cosh(\xi,\alpha)+3\sinh(\xi,\alpha)+4r}, \tag{43}$$

$$u_3 = \frac{\sqrt{\sigma(r^2+M)}\,\mathrm{csch}(\xi,\alpha)+\sqrt{\sigma}\,\coth(\xi,\alpha)}{r\,\mathrm{csch}(\xi,\alpha)+1}, \tag{44}$$

where $\xi = k\,x + i\,k\,y + \omega t,\ \omega^\alpha = -2b_1\tau$. When $\alpha = 1$ equation (38) reduced to the well known biological population model

$$u_t = u_{xx} + u_{yy} + \tau(u^2-\sigma), \tag{45}$$

The solutions (42) - (44) take the form of the following solutions of the biological population model

$$u_{1bio} = \frac{\sqrt{\sigma(r^2-1)}\,\mathrm{sech}(\xi) + \sqrt{\sigma}\,\tanh(\xi)}{r\,\mathrm{sech}(\xi)+1}, \tag{46}$$

$$u_{2bio} = \frac{4\sqrt{\sigma(r^2-1)} + \sqrt{\sigma}\left[5\sinh(\xi)+3\cosh(\xi)\right]}{5\cosh(\xi)+3\sinh(\xi)+4r}, \tag{47}$$

$$u_{3bio} = \frac{\sqrt{\sigma(r^2+1)}\,\mathrm{csch}(\xi) + \sqrt{\sigma}\,\coth(\xi)}{r\,\mathrm{csch}(\xi)+1}, \tag{48}$$

**Remark 1:** The FDEs describe discontinuous media and the fractional order is equivalent to its fractional dimensions. If we consider a plane with fractal structure, the shortest path between two points is not a line and we have [48, 49]

$$D_x^{\alpha} f(x) \cong \Gamma(\alpha+1) D_x f(x). \tag{49}$$

Prof. He et al in [50] modified the chain rule to the formulae

$$D_x^{\alpha} f[g(x)] = \sigma_x f_g^{'}[g(x)] D_x^{\alpha} g(x), \tag{50}$$

where $\sigma_x$ is called the fractal index which is usually determined in terms of gamma functions [49, 50]. In this paper, we use the modified formulae of chain rule that read

$$D_x^{\alpha} f[g(x)] = \sigma_x D_g^{\alpha} f[g(x)](g_x^{'})^{\alpha}. \tag{51}$$

To compute the fractal index $\sigma_x$ where the obtained results depend on $\xi$ and $E_{\alpha}(\xi^{\alpha})$.

Let $f(t) = E_{\alpha}(t^{\alpha}) = \sum_{k=0}^{\infty} \frac{t^{\alpha k}}{\Gamma(1+\alpha k)}$ and $g(x) = x,$ then the derivative of the right hand side is

$$D_x^{\alpha} f[g(x)] = D_x^{\alpha} E_{\alpha}(x^{\alpha}) = E_{\alpha}(x^{\alpha}). \tag{52}$$

The left hand side is

$$\sigma_x D_g^{\alpha} f[g(x)](g_x^{'})^{\alpha} = \sigma_x E_{\alpha}(x^{\alpha})(1)^{\alpha} = \sigma_x E_{\alpha}(x^{\alpha}). \tag{53}$$

So the fractal index is equal to one $(\sigma_x = 1)$. Counter examples are introduced in appendix B to show how to compute the fractal index.

## 4. Conclusion

In this paper, to construct exact analytical solutions of nonlinear FDEs, the fractional projective Riccati expansion method is presented. The space-time fractional Burgers equation, the space-time fractional mKdV equation and Time fractional biological population model are chosen to demonstrate the power of the method. To the best of our knowledge, some of the solutions obtained in this research paper have not been reported in literature. Mathematical packages can be used to perform more complicated and tedious algebraic calculations. The projective Riccati expansion method can be applied to other nonlinear FDEs. The graph of the solutions of the Burgers equation shows that the fractional order changes both the height and the width of the waves. This means that the fractional order modulate the shape of the wave. The obtained results are invariant under the consideration of He et al in [50] since the fractal index is equal to one.

**Appendix A**

The product of two power series are given by

$$\left(\sum_{n=0}^{\infty} a_n x^n\right)\left(\sum_{n=0}^{\infty} b_n x^n\right) = \sum_{n=0}^{\infty} c_n x^n, \tag{A-1}$$

where $c_n = \sum_{k=0}^{\infty} a_k b_{n-k}$ . If $n$ is a natural number, then

$$\left(\sum_{k=0}^{\infty} a_k x^k\right)^n = \sum_{k=0}^{\infty} c_k x^k, \tag{A-2}$$

where $c_o = a_0^n,\ \ c_m = \dfrac{1}{m\,a_0}\sum_{k=1}^{\infty}(kn-m+k)\,a_k c_{m-k}$ . For simplicity, we suppose that

$$E_\alpha(x^\alpha)E_\alpha(-x^\alpha) = \left(\sum_{k=0}^{\infty}\frac{x^{\alpha k}}{\Gamma(1+\alpha k)}\right)\left(\sum_{k=0}^{\infty}\frac{(-x^\alpha)^k}{\Gamma(1+\alpha k)}\right) = M. \tag{A-3}$$

From the definition of $\cosh(x,\alpha)$ and $\sinh(x,\alpha)$, we can get the following inequality

$$\begin{aligned}\cosh^2(x,\alpha)-\sinh^2(x,\alpha) &= \frac{1}{4}[E_\alpha(x^\alpha)^2 + 2E_\alpha(x^\alpha)E_\alpha(-x^\alpha) + E_\alpha(-x^\alpha)^2 - E_\alpha(x^\alpha)^2 \\ &\quad + 2E_\alpha(x^\alpha)E_\alpha(-x^\alpha) - E_\alpha(-x^\alpha)^2] = E_\alpha(x^\alpha)E_\alpha(-x^\alpha) = M.\end{aligned} \tag{A-4}$$

Dividing by $\cosh^2(x,\alpha)$ and $\sinh^2(x,\alpha)$, we have

$$1-\tanh^2(x,\alpha) = M\,\mathrm{sech}^2(x,\alpha), \tag{A-5}$$

$$\coth^2(x,\alpha) - 1 = M\,\mathrm{csch}^2(x,\alpha). \tag{A-6}$$

Similarly, we suppose that

$$E_\alpha(i\,x^\alpha)E_\alpha(-i\,x^\alpha)=\left(\sum_{k=0}^{\infty}\frac{(i\,x^\alpha)^k}{\Gamma(1+\alpha k)}\right)\left(\sum_{k=0}^{\infty}\frac{(-i\,x^\alpha)^k}{\Gamma(1+\alpha k)}\right)=\tilde{M}. \quad \text{(A-7)}$$

$$\cos^2(x,\alpha)+\sin^2(x,\alpha)=E_\alpha(i\,x^\alpha)E_\alpha(-i\,x^\alpha)=\tilde{M}. \quad \text{(A-8)}$$

$$1+\tan^2(x,\alpha)=\tilde{M}\sec^2(x,\alpha), \quad \text{(A-9)}$$

$$\cot^2(x,\alpha)+1=\tilde{M}\csc^2(x,\alpha). \quad \text{(A-10)}$$

The fractional derivatives of the Mittag-Leffler function take the form

$$\begin{aligned} D_x^\alpha E_\alpha(x^\alpha) &= \sum_{k=0}^{\infty}\frac{D_x^\alpha\, x^{\alpha k}}{\Gamma(1+\alpha k)}=\sum_{k=1}^{\infty}\frac{\Gamma(1+\alpha k)\; x^{\alpha k-\alpha}}{\Gamma(1+\alpha k)\Gamma(\alpha k+1-\alpha)}=\sum_{k=1}^{\infty}\frac{x^{\alpha(k-1)}}{\Gamma(\alpha(k-1)+1)} \\ &=\sum_{s=0}^{\infty}\frac{x^{\alpha s}}{\Gamma(\alpha s+1)}=E_\alpha(x^\alpha), \end{aligned} \quad \text{(A-11)}$$

$$\begin{aligned} D_x^\alpha E_\alpha(-x^\alpha) &= \sum_{k=0}^{\infty}\frac{(-1)^k D_x^\alpha\, x^{\alpha k}}{\Gamma(1+\alpha k)}=\sum_{k=1}^{\infty}\frac{(-1)^k\Gamma(1+\alpha k)\; x^{\alpha k-\alpha}}{\Gamma(1+\alpha k)\Gamma(\alpha k+1-\alpha)}=\sum_{k=1}^{\infty}\frac{(-1)^k\; x^{\alpha(k-1)}}{\Gamma(\alpha(k-1)+1)} \\ &=\sum_{s=-}^{\infty}\frac{(-1)^{s+1}\; x^{\alpha s}}{\Gamma(\alpha s+1)}=-E_\alpha(-x^\alpha), \end{aligned} \quad \text{(A-12)}$$

$$D_x^\alpha E_\alpha(i\,x^\alpha)=i\,E_\alpha(i\,x^\alpha), \quad \text{(A-13)}$$

$$D_x^\alpha E_\alpha(-i\,x^\alpha)=-i\,E_\alpha(-i\,x^\alpha). \quad \text{(A-14)}$$

From equations (A-11) and (A-12), we can get the derivatives of the generalized hyperbolic functions

$$D_x^\alpha[\sinh(x,\alpha)]=\frac{D_x^\alpha[E_\alpha(x^\alpha)]-D_x^\alpha[E_\alpha(-x^\alpha)]}{2}=\frac{E_\alpha(x^\alpha)+E_\alpha(-x^\alpha)}{2}=\cosh(x,\alpha), \quad \text{(A-15)}$$

$$D_x^\alpha[\cosh(x,\alpha)]=\frac{D_x^\alpha[E_\alpha(x^\alpha)]+D_x^\alpha[E_\alpha(-x^\alpha)]}{2}=\frac{E_\alpha(x^\alpha)-E_\alpha(-x^\alpha)}{2}=\sinh(x,\alpha), \quad \text{(A-16)}$$

$$D_x^\alpha[\sin(x,\alpha)]=\frac{D_x^\alpha[E_\alpha(i\,x^\alpha)]-D_x^\alpha[E_\alpha(-i\,x^\alpha)]}{2i}=\frac{E_\alpha(ix^\alpha)+E_\alpha(-ix^\alpha)}{2}=\cos(x,\alpha), \quad \text{(A-17)}$$

$$D_x^\alpha[\cos(x,\alpha)]=\frac{D_x^\alpha[E_\alpha(ix^\alpha)]+D_x^\alpha[E_\alpha(-ix^\alpha)]}{2}=-\frac{E_\alpha(ix^\alpha)-E_\alpha(-ix^\alpha)}{2i}=-\sin(x,\alpha), \quad \text{(A-18)}$$

By using these inequalities, we proof the three cases.

Case 1: when $\mu=-M$ , we have $f_1=\dfrac{\operatorname{sech}(\xi,\alpha)}{r\operatorname{sech}(\xi,\alpha)+1}$, $\quad g_1=\dfrac{\tanh(\xi,\alpha)}{r\operatorname{sech}(\xi,\alpha)+1}$.

Firstly, we proof the relation $D^\alpha f=-f\,g$,

$$D^{\alpha} f_1 = D^{\alpha}\left[\frac{\operatorname{sech}(\xi,\alpha)}{r\operatorname{sech}(\xi,\alpha)+1}\right]$$

$$= \frac{-\operatorname{sech}(\xi,\alpha)\tanh(\xi,\alpha)\left(r\operatorname{sech}(\xi,\alpha)+1\right)+r(\operatorname{sech}(\xi,\alpha))^2\tanh(\xi,\alpha)}{\left(r\operatorname{sech}(\xi,\alpha)+1\right)^2} \tag{A-19}$$

$$= \frac{-\operatorname{sech}(\xi,\alpha)\tanh(\xi,\alpha)}{\left(r\operatorname{sech}(\xi,\alpha)+1\right)^2} = (-1)\left(\frac{\operatorname{sech}(\xi,\alpha)}{r\operatorname{sech}(\xi,\alpha)+1}\right)\left(\frac{\tanh(\xi,\alpha)}{r\operatorname{sech}(\xi,\alpha)+1}\right) = -f_1\, g_1.$$

Secondly, we proof the relation $D^{\alpha} g = 1 - g^2 - r\, f$,

$$\text{R.H.S} = D^{\alpha} g_1 = D^{\alpha}\left(\frac{\tanh(\xi,\alpha)}{r\operatorname{sech}(\xi,\alpha)+1}\right)$$

$$= \frac{M(\operatorname{sech}(\xi,\alpha))^2(r\operatorname{sech}(\xi,\alpha)+1)+r\operatorname{sech}(\xi,\alpha)(\tanh(\xi,\alpha))^2}{\left(r\operatorname{sech}(\xi,\alpha)+1\right)^2}$$

$$= \frac{M(\operatorname{sech}(\xi,\alpha))^2(r\operatorname{sech}(\xi,\alpha)+1)+r\operatorname{sech}(\xi,\alpha)(1-M(\operatorname{sech}(\xi,\alpha))^2)}{\left(r\operatorname{sech}(\xi,\alpha)+1\right)^2} \tag{A-20}$$

$$= \frac{M(\operatorname{sech}(\xi,\alpha))^2 + r\operatorname{sech}(\xi,\alpha)}{\left(r\operatorname{sech}(\xi,\alpha)+1\right)^2} = \frac{1-(\tanh(\xi,\alpha))^2 + r\operatorname{sech}(\xi,\alpha)}{\left(r\operatorname{sech}(\xi,\alpha)+1\right)^2}.$$

$$\text{L.H.S} = 1 - g_1^2 - r\, f_1 = 1 - \left(\frac{\tanh(\xi,\alpha)}{r\operatorname{sech}(\xi,\alpha)+1}\right)^2 - r\left(\frac{\operatorname{sech}(\xi,\alpha)}{r\operatorname{sech}(\xi,\alpha)+1}\right)$$

$$= \frac{\left(r\operatorname{sech}(\xi,\alpha)+1\right)^2 - (\tanh(\xi,\alpha))^2 - r\operatorname{sech}(\xi,\alpha)\left(r\operatorname{sech}(\xi,\alpha)+1\right)}{\left(r\operatorname{sech}(\xi,\alpha)+1\right)^2} \tag{A-21}$$

$$= \frac{1-(\tanh(\xi,\alpha))^2 + r\operatorname{sech}(\xi,\alpha)}{\left(r\operatorname{sech}(\xi,\alpha)+1\right)^2}.$$

Finally, we proof that $g^2 = 1 - 2r\, f + (r^2+\mu)\, f^2$,

$$1 - 2r\, f + (r^2+\mu)\, f^2 = 1 - 2r\left(\frac{\operatorname{sech}(\xi,\alpha)}{r\operatorname{sech}(\xi,\alpha)+1}\right) + (r^2 - M)\left(\frac{\operatorname{sech}(\xi,\alpha)}{r\operatorname{sech}(\xi,\alpha)+1}\right)^2$$

$$= \frac{\left(r\operatorname{sech}(\xi,\alpha)+1\right)^2 - 2r\operatorname{sech}(\xi,\alpha)\left(r\operatorname{sech}(\xi,\alpha)+1\right) + (r^2 - M)(\operatorname{sech}(\xi,\alpha))^2}{\left(r\operatorname{sech}(\xi,\alpha)+1\right)^2} \tag{A-22}$$

$$= \frac{1 - M(\operatorname{sech}(\xi,\alpha))^2}{\left(r\operatorname{sech}(\xi,\alpha)+1\right)^2} = \left(\frac{\tanh(\xi,\alpha)}{r\operatorname{sech}(\xi,\alpha)+1}\right)^2 = g^2,$$

this complete the proof. By the same manner, the two cases can be proved.

Appendix B

To explain how to compute the fractal index from the derivative of multiplication rule of two functions four examples are introduced.

Example 1: Suppose that $f$, $g$ are two $\alpha$-differentiable functions that is

$$f = x^{\beta}, \quad g = x^{\gamma}, \qquad 0 < \beta < 1,\ 0 < \gamma < 1. \tag{B-1}$$

We compute the fractal index by using the modified multiplication rule of two functions

$$D_x^{\alpha}[f(x)g(x)] = \sigma_x [g(x) D_x^{\alpha} f(x) + f(x) D_x^{\alpha} g(x)]. \tag{B-2}$$

The derivative of the right hand side is

$$D_x^{\alpha}[f(x)g(x)] = D_x^{\alpha}(x^{\beta+\gamma}) = \frac{\Gamma(\beta+\gamma+1)}{\Gamma(\beta+\gamma+1-\alpha)} x^{\beta+\gamma-\alpha}. \tag{B-3}$$

The left hand side is

$$\begin{aligned} \sigma_x [g(x) D_x^{\alpha} f(x) + f(x) D_x^{\alpha} g(x)] &= \sigma_x [x^{\beta} D_x^{\alpha}(x^{\gamma}) + x^{\gamma} D_x^{\alpha}(x^{\beta})] \\ &= \sigma_x \left[ \frac{\Gamma(\gamma+1)}{\Gamma(\gamma+1-\alpha)} + \frac{\Gamma(\beta+1)}{\Gamma(\beta+1-\alpha)} \right] x^{\beta+\gamma-\alpha}. \end{aligned} \tag{B-3}$$

Therefore, the fractal index $\sigma_x$ is equal to

$$\sigma_x = \frac{\Gamma(\beta+\gamma+1)}{\Gamma(\beta+\gamma+1-\alpha)} \left[ \frac{\Gamma(\beta+1-\alpha)\Gamma(\gamma+1-\alpha)}{\Gamma(\gamma+1-\alpha)\Gamma(\beta+1) + \Gamma(\beta+1-\alpha)\Gamma(\gamma+1)} \right]. \tag{B-4}$$

Example 2: Suppose that $f$, $g$ are two $\alpha$-differentiable functions that is

$$f = g = x^{\alpha}, \qquad 0 < \alpha < 1. \tag{B-5}$$

The fractal index $\sigma_x$ is equal to

$$\sigma_x = \frac{\Gamma(2\alpha+1)}{2[\Gamma(\alpha+1)]^2}. \tag{B-6}$$

Example 3: Suppose that one of the functions $f$, $g$ is $\alpha$-differentiable function and the second is differentiable function that is

$$f = x, \qquad g = x^{\beta}, \qquad 0 < \beta < 1. \tag{B-7}$$

The derivative of the right hand side is

$$D_x^{\alpha}[f(x)g(x)] = D_x^{\alpha}(x^{\beta+1}) = \frac{\Gamma(\beta+2)}{\Gamma(\beta+2-\alpha)} x^{\beta+1-\alpha}. \tag{B-8}$$

The left hand side is

$$\sigma_x[g(x)D_x^\alpha f(x)+f(x)D_x^\alpha g(x)]=\sigma_x[x^\beta D_x^\alpha(x)+xD_x^\alpha(x^\beta)]$$
$$=\sigma_x\left[\frac{\Gamma(2)}{\Gamma(2-\alpha)}+\frac{\Gamma(\beta+1)}{\Gamma(\beta+1-\alpha)}\right]x^{\beta+1-\alpha}. \tag{B-9}$$

Therefore, the fractal index $\sigma_x$ is equal to

$$\sigma_x=\frac{\Gamma(\beta+2)}{\Gamma(\beta+2-\alpha)}\left[\frac{\Gamma(\beta+1-\alpha)\Gamma(2-\alpha)}{\Gamma(2-\alpha)\Gamma(\beta+1)+\Gamma(2-\alpha)\Gamma(2)}\right]. \tag{B-10}$$

Example 4: Suppose that the two functions $f$, $g$ are differentiable functions that is

$$f=g=x. \tag{B-11}$$

The derivative of the right hand side is

$$D_x^\alpha[f(x)g(x)]=D_x^\alpha(x^2)=\frac{\Gamma(3)}{\Gamma(3-\alpha)}x^{2-\alpha}. \tag{B-12}$$

The lift hand side is

$$\sigma_x[g(x)D_x^\alpha f(x)+f(x)D_x^\alpha g(x)]=2\sigma_x xD_x^\alpha(x)=\frac{2\sigma_x\Gamma(2)}{\Gamma(2-\alpha)}x^{2-\alpha}. \tag{B-13}$$

So the fractal index $\sigma_x$ is equal to

$$\sigma_x=\frac{\Gamma(2-\alpha)}{\Gamma(3-\alpha)}. \tag{B-14}$$

To explain how to compute the fractal index from the chain rule given by equation (4) two examples are introduced.

Example 5: Suppose that $f$ is a differentiable function and $g$ is an $\alpha$-differentiable function that is

$$f=t^n, \qquad g=x^\beta, \qquad 0<\beta<1. \tag{B-15}$$

We compute the fractal index by using the modified chain rule which given by

$$D_x^\alpha f[g(x)]=\sigma_x\, f_g^{'}[g(x)]D_x^\alpha g(x). \tag{B-16}$$

The derivative of the right hand side is

$$D_x^\alpha f(g(x))=D_x^\alpha(x^{n\beta})=\frac{\Gamma(n\beta+1)}{\Gamma(n\beta+1-\alpha)}x^{n\beta-\alpha}. \tag{B-17}$$

The left hand side is

$$\sigma_x\ f_g^{'}[g(x)]D_x^{\alpha}g(x)=\frac{\sigma_x n(x^{\beta})^{n-1}\Gamma(\beta+1)x^{\beta-\alpha}}{\Gamma(\beta+1-\alpha)}$$
$$=\frac{\sigma_x n\Gamma(\beta+1)}{\Gamma(\beta+1-\alpha)}x^{n\beta-\alpha} \tag{B-18}$$

The fractal index $\sigma_x$ is equal to

$$\sigma_x=\frac{\Gamma(\beta+1-\alpha)\Gamma(n\beta+1)}{n\Gamma(\beta+1)\Gamma(n\beta+1-\alpha)}. \tag{B-19}$$

Example 6: Suppose $f$ is a differentiable function and $g$ is an $\alpha$-differentiable function that is

$$f=t^n, \qquad g=x^{\alpha}, \qquad 0<\alpha<1. \tag{B-20}$$

The fractal index $\sigma_x$ is equal to

$$\sigma_x=\frac{\Gamma(n\alpha+1)}{n\Gamma(\alpha+1)\Gamma(n\alpha+1-\alpha)}. \tag{B-21}$$

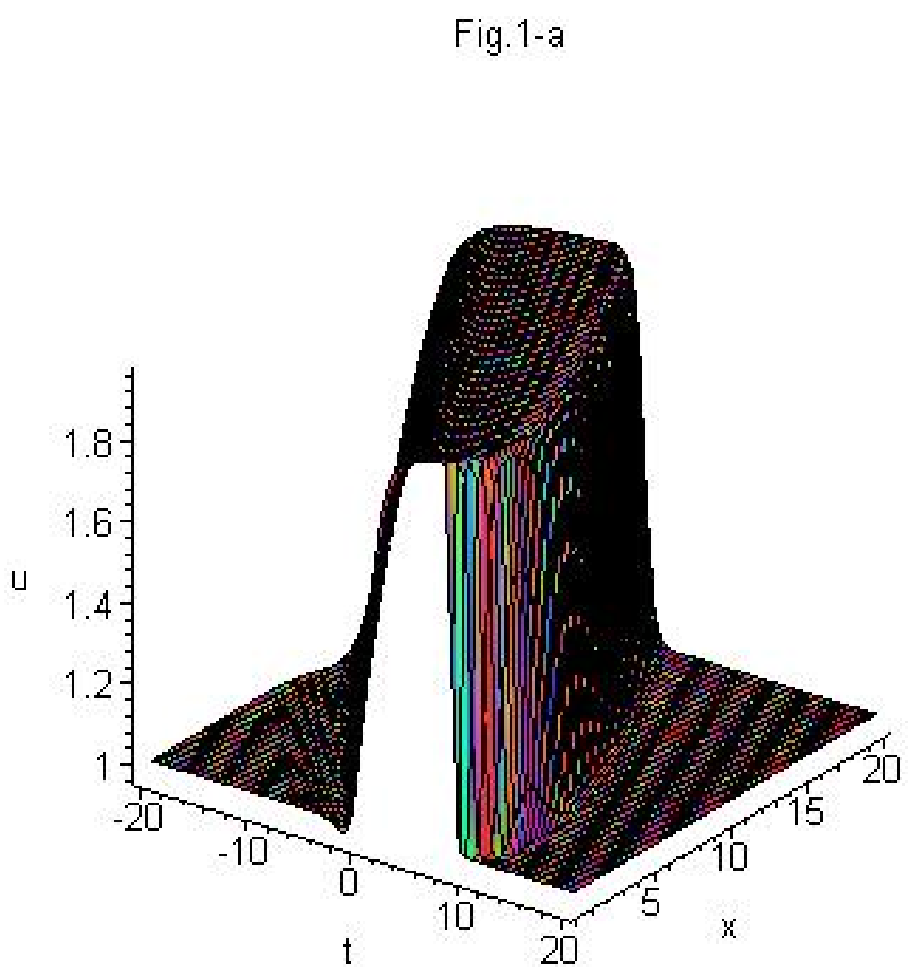

**Figure (1-a):** Solution (20) of Burgers equation describes bell-shaped with $k = \tau = \omega = 1,\ r = 2$ when $\alpha = 0.75$ ..

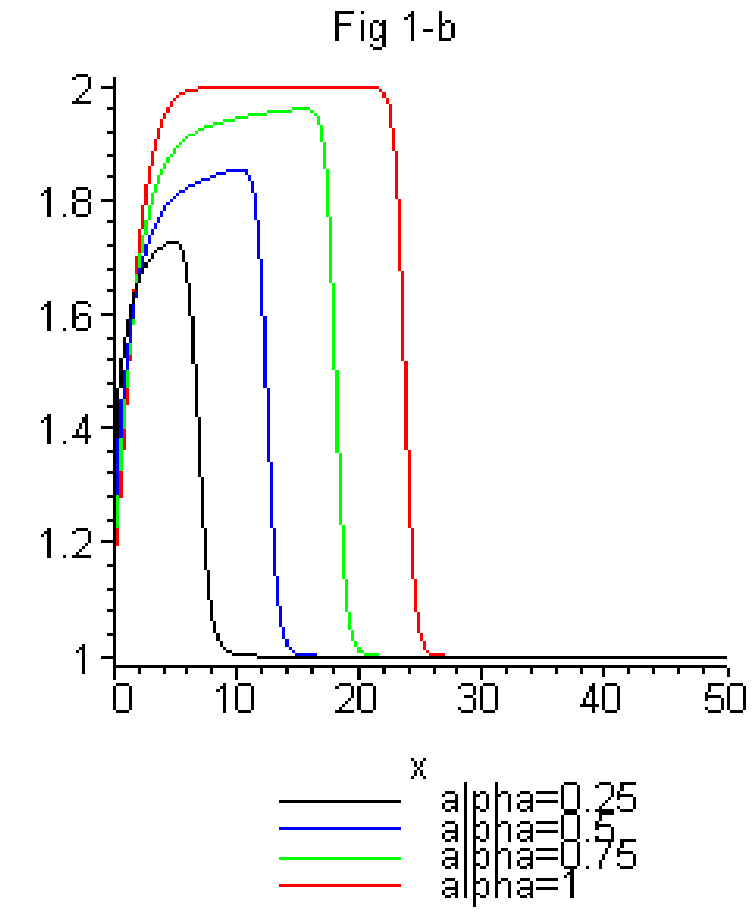

Figure (1- b): the 2-dimensional behavior of solution (20) at different values of $\alpha$

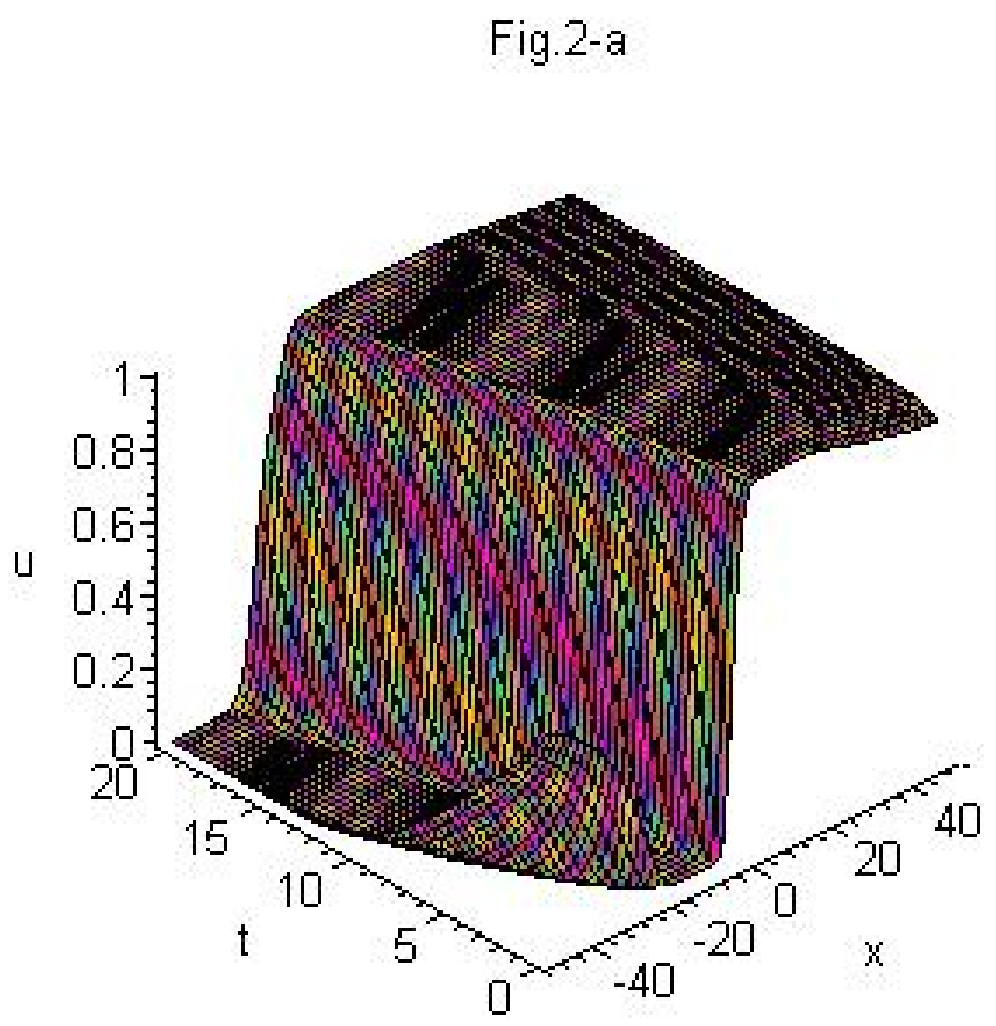

**Figure (2- a):** Solution (21) of Burgers equation describes kink-shaped

With $k = 2,\ \tau = \omega = 1,\ r = \sqrt{5}$ when $\alpha = 0.8$ .

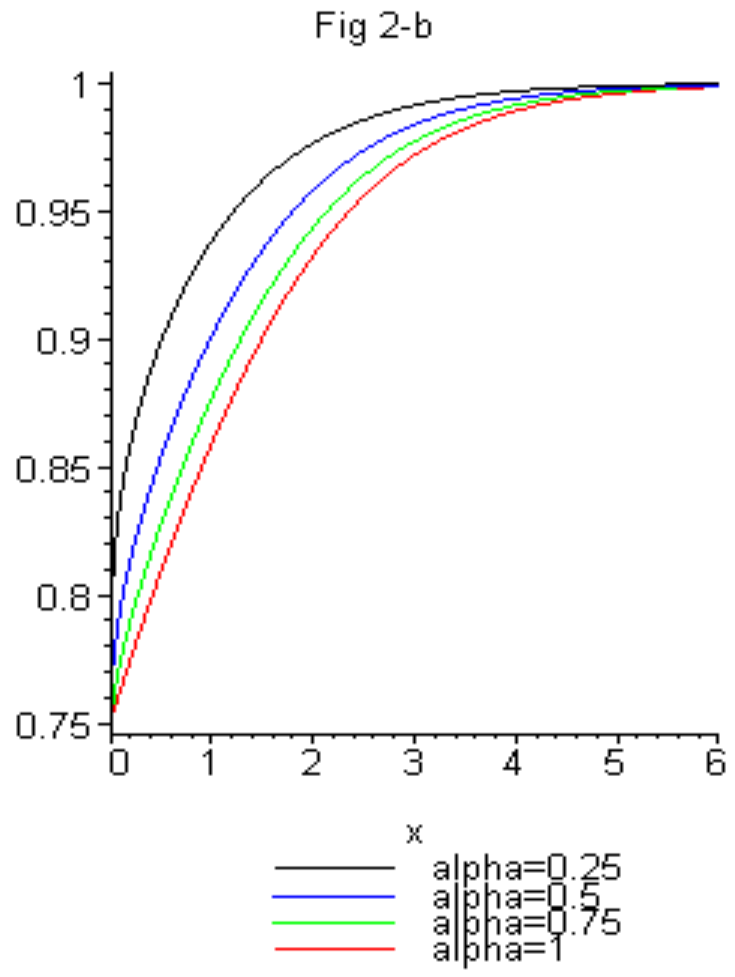

Figure (1- b): the 2-dimensional behavior of solution (21) at different values of $\alpha$